\RequirePackage{ifpdf}
\documentclass[letterpaper]{JHEP3}
\usepackage{array}
\usepackage{float}
\usepackage{nicefrac}
\usepackage{amsmath,epsfig}
\usepackage{amssymb,amsfonts,mathrsfs}
\usepackage{color}
\usepackage{subfigure}
\usepackage{skak}

\usepackage{graphics}
\usepackage{epsfig}
\usepackage{amsfonts}

\renewcommand{\Im}{\mbox{Im}}

\newcommand{\PTerm}[2]{\big\{{\textstyle\genfrac{}{}{0pt}{}{#1}{#2}}\big\}}
\newcommand{\atopfrac}[2]{\genfrac{}{}{0pt}{}{#1}{#2}}
\newcommand{\sfrac}[2]{{\textstyle\frac{#1}{#2}}}
\newcommand{\half}{\sfrac{1}{2}}

\newcommand{\bigbrk}[1]{\bigl(#1\bigr)}

\newcommand{\nln}{\nonumber\\}
\newcommand{\nl}{\nonumber\\&&\mathord{}}

\newcommand{\earel}[1]{\mathrel{}&#1&\mathrel{}}
\newcommand{\eq}{\earel{=}}

\newenvironment{myeqnarray*}{\arraycolsep0pt\begin{eqnarray*}}{\end{eqnarray*}\ignorespacesafterend}

\newcommand{\nn}{\nonumber}

\global \long \def \NN{ \mathcal{N}}

\global \long \def \Im{ \mathcal{I}}
\global \long \def \Jm{ \mathcal{J}}
\global \long \def \Nm{ \mathcal{N}}
\global \long \def \Qm{ \mathcal{Q}}
\global \long \def \Sm{ \mathcal{S}}
\global \long \def \Lm{ \mathcal{L}}
\global \long \def \Rm{ \mathcal{R}}
\global \long \def \Pm{ \mathcal{P}}
\global \long \def \Km{ \mathcal{K}}
\global \long \def \Dm{ \mathcal{D}}
\global \long \def \Fm{ \mathcal{F}}
\global \long \def \Um{ \mathcal{U}}
\global \long \def \Hm{ \mathcal{H}}

\global \long \def \lra{\leftrightarrow}
\global \long \def \ph{\phantom}

\def\gb{\beta}

\def\half{{\frac12}}

\def\IC{\relax\hbox{$\inbar\kern-.3em{\rm C}$}}

\def\IC{{\bf C}}

\def\bea{\begin{eqnarray}}
\def\eea{\end{eqnarray}}
\def\be{\begin{equation}}
\def\ee{\end{equation}}
\def\ba{\begin{align}}
\def\ea{\end{align}}
\def\bse{\begin{subequations}}
\def\ese{\end{subequations}}
\def\1F1{{}_1\!F_1}
\def\2F0{{}_2\!F_0}

\def\nn{\nonumber}

\def\a{\alpha}
\def\b{\beta}
\def\ad{\dot{\alpha}}
\def\bd{\dot{\beta}}

\def\h3{$\textrm{H}_3^+$}

\def\IC{{\mathbb C}}

\def\lbldef#1#2{\expandafter\gdef\csname #1\endcsname {#2}}

\def\href#1#2{#2}

\newcommand{\beq}{\begin{equation}}
\newcommand{\eeq}{\end{equation}}
\newcommand{\ber}{\begin{eqnarray}}
\newcommand{\eer}{\end{eqnarray}}

\def\be{\begin{eqnarray}}
\def\ee{\end{eqnarray}}
\def\SS{\scriptscriptstyle}

\newcommand{\f}{{\phi}}
\newcommand{\g}{\gamma}

\providecommand{\tabularnewline}{\\}

\newcommand{\rep}[1]{\mathbf{#1}}

\newcommand{\ket}[1]{|#1\rangle}

\newcommand{\MMM}[2]{{\arraycolsep0pt\begin{array}[b]{c}\makebox[0cm]{$\atopfrac{#2}{\downarrow}$}\vspace{-0.5cm}\\#1\end{array}}}

\ifx\genfrac\sdflkaj
\newcommand{\atopfrac}[2]{{{#1}\above0pt{#2}}}
\else
\fi

\def\({\left(}
\def\){\right)}
\def\[{\left[}
\def\]{\right]}
\def\<{\langle}
\def\>{\rangle}

\def\MM{{\cal M}}

\def\SPsi{\Psi_\rep{1}}
\def\TPsi{\Psi_\rep{3}}
\def\nPsi{\Psi_{n}}
\def\MPsi{\Psi_\MM}

\def\SS{{S_\rep{1}}}
\def\TS{{S_\rep{3}}}
\def\nS{S_{n}}
\def\MS{{S_\MM}}

\title{\begin{center} On the Integrability of Planar \\ ${\cal N}=2$ Superconformal Gauge Theories
\end{center}}
\preprint{YITP-SB-12-38}
\author{Abhijit Gadde$^{a, b}$\footnote{Email: abhijit@caltech.edu}\;,
$\,$ Pedro Liendo$^{b}$\footnote{Email: pedro.liendo@stonybrook.edu}\;,
$\,$ Leonardo Rastelli$^{b}$\footnote{Email: leonardo.rastelli@stonybrook.edu}\;,
$\,$ and Wenbin Yan$^{a, b}$\footnote{Email: wbyan@theory.caltech.edu}
\\
\\
\it $^a$ California Institute of Technology,\\
\it Pasadena, CA 91125, USA 
\\
\\
\it $^b$ C.N. Yang Institute for Theoretical Physics,\\
\it Stony Brook University, \\
\it Stony Brook, NY 11794-3840, USA
}

\bigskip

\abstract{We study 
the integrability properties of planar ${\cal N}=2$ superconformal field theories in four dimensions.
We show that the  spin chain associated to the planar dilation operator  of  ${\cal N}=2$ superconformal  QCD fails to be integrable at two loops.
In our analysis we focus on a closed $SU(2|1)$ sector, whose two-loop spin chain we fix by symmetry arguments
(up to a few undetermined coefficients). It turns out that the Yang-Baxter equation for magnon scattering is not satisfied in this sector.
On the other hand, we suggest that the closed $SU(2,1|2)$ sector, which exists in any ${\cal N}=2$ superconformal gauge theory,
may be integrable to all loops. We summarize the known results in the literature that are consistent with this conjecture.
}

\begin{document}

\section{Introduction and Summary}

There is by now overwhelming evidence that
planar ${\cal N}=4$ super-Yang Mills theory is a completely
integrable model (see \cite{Beisert:2010jr} for a comprehensive review).
To which extent  integrability survives in less symmetric (and more realistic)  
gauge  theories is
  an important question, both because  integrability is a very  useful computational tool,
  and because exploring  a larger set of examples should shed light on its
  conceptual origin,   which is still mysterious. 
 In fact, the first instances of integrability in a four-dimensional gauge theory
 were found in  QCD itself \cite{Lipatov:1993yb,Faddeev:1994zg,Korchemsky:1994um,Braun:1998id,Braun:1999te,Belitsky:1999bf}. However, with
 hindsight,
  the integrability properties of large $N_c$ QCD discovered  so far
  can be understood as being ``inherited'' from the maximally supersymmetric theory. 
For example, a large sector   of QCD composite operators  has identical  one-loop
renormalization  
 as the analogous sector in ${\cal N}=4$ SYM.\footnote{The maximal one-loop integrable sector in QCD is the $SU(2,2)$ sector
  described in \cite{Beisert:2004fv}. It contains the $SL(2, \mathbb{R})$ sector of  maximal helicity ``quasipartonic'' lightcone operators.
 In this latter sector,  the planar dilation operator has been shown to coincide with that of ${\cal N}=4$ SYM also at two loops \cite{Belitsky:2004yg,Belitsky:2004sc}, up to 
  overall factors that capture the non-vanishing beta function and the non-universal regulator dependence.}
 At higher loops, the analysis of the QCD dilation operator
   is complicated by the breaking of conformal invariance and by the (non-universal) dependence
  on the regulator. A parallel story holds for
  ${\cal N}=1$ and ${\cal N}=2$ supersymmetric Yang-Mills theories in the usual 't Hooft limit (large $N_c$, fixed $N_f$), see \cite{Belitsky:2004yg,Belitsky:2004sc,Belitsky:2004sf,Belitsky:2005bu,Korchemsky:2010kj} and references therein.
  This motivates us to explore integrability  in the cleaner theoretical
  laboratory of theories that remain exactly conformal at the quantum level.
  The main question
    one would like to answer is whether integrability in less symmetric conformal gauge theories is always an ``accidental'' remnant of  the ${\cal N}=4$ integrability
    (and under which conditions do such accidents occur),
    or whether genuinely new structures are also possible.
    
  A large class of four-dimensional conformal  theories
  are the ${\cal N}=2$ supersymmetric theories with vanishing one-loop beta function.
A  well-known non-renormalization theorem guarantees that the beta function remains zero in the full quantum theory.
 Perhaps the simplest example (beyond ${\cal N}=4$ SYM itself)
  is ${\cal N}=2$ superconformal  QCD (SCQCD), the theory with gauge group $SU(N_c)$ and $2 N_c$ fundamental
  hypermultiplets. Integrability is {\it at best} expected in the planar   
  Veneziano limit of large $N_c$ {\it and} large $N_f \equiv 2 N_c$, with fixed 't Hooft coupling $\lambda = g_{YM}^2 N_c$.

The dilation operator of planar SCQCD defines, as usual, the Hamiltonian of a spin chain.\footnote{See {\it e.g.} \cite{Minahan:2002ve,Beisert:2003tq,Beisert:2003jj,Beisert:2003ys,Zwiebel:2005er,Sieg:2010tz,Zwiebel:2011bx} and the reviews \cite{Minahan:2010js,Sieg:2010jt} for a very partial list of references on the evaluation of the dilation operator in ${\cal N}=4$ SYM. See also \cite{Zoubos:2010kh} for a review of the dilation operator in deformations of $\Nm$=4 SYM.} 
We review its symmetry structure in Section 2.
Closed chains correspond to flavor singlet gauge-invariant operators of the schematic form \cite{Gadde:2009dj,Gadde:2010zi}
${\rm Tr} \left(  \varphi^{k_1} {\cal M}^{k_2} \varphi^{k_3} {\cal M}^{k_4} \dots \right)
$. Here $\varphi$ denotes any of the color-adjoint elementary ``letters'', for example 
 $\varphi  =  ({\cal D}^n \lambda)^{a}_{\ph{a} b}$, where ${\cal D}$ is a gauge-covariant derivative, $\lambda$ a gaugino field,
 and $a, b = 1, \dots N_c$ color indices. The symbol $\MM$ stands for any of the gauge-adjoint
 composite dimers obtained by the flavor contraction of a fundamental and a antifundamental letter,
 for example $\MM^a_{\ph{b}b}  = Q^{a i} \bar Q_{bi}$, where $Q$ is the squark field and $i =1, \dots N_f$ a flavor index.
One can also consider open chains with open flavor indices at the endpoints.

The one-loop Hamiltonian of ${\cal N}=2$ SCQCD was evaluated in the sector of composite operators  made of elementary scalar fields in \cite{Gadde:2010zi},
and for the full theory in \cite{Liendo:2011xb}.  The question of its integrability is still not completely 
settled.
Despite some early intriguing hints \cite{Gadde:2010zi},
the spectrum of anomalous dimensions  does not exhibit \cite{Liendo:2011wc} the systematic pairing of opposite-parity eigenvalues
 that is one of the hallmarks of integrability \cite{Beisert:2003tq,Beisert:2003ys,Zwiebel:2005er,Beisert:2004ry}. It is often easy to {\it disprove} integrability
 by setting up a position-space Bethe ansatz and showing that  the $n$-body magnon S-matrix does {\it not} factorize.
 In our case, this is not straightforward  because the S-matrix of external  {\it dimeric} magnons
 (${\cal M}$'s moving on the chain) is hard to calculate. On the other hand, the S-matrix of the {\it elementary} (single-letter) magnons
 is unaffected at one loop by the presence of the dimers, and trivially coincides with a restriction 
 of the ${\cal N}=4$ S-matrix -- an instance of ``accidental'' one-loop integrability inherited from ${\cal N}=4$ SYM.
 
 As it turns out, it is easier to test integrability at two loops. In Section 3 we consider  a simple closed $SU(2|1)$ sector,
 and fix its two-loop Hamiltonian using symmetry, up to a few undetermined parameters. 
 This sector is particularly interesting because it is structurally different from any 
  subsector of ${\cal N}=4$ SYM, as the dimers play a crucial role.
 The asymptotic excitations on the $SU(2|1)$ chain are gauginos $\lambda_{\alpha}$, where $\alpha$ is an $SU(2)$ Lorentz index.
 In Section 4 we evaluate
 their two-body scattering matrix
 and find that it fails to satisfy
 the Yang-Baxter equation, 
 which conclusively shows that the  Hamiltonian of ${\cal N}=2$ SCQCD is {\it not} completely integrable at higher loops. 
 This would have required a novel integrability structure (not present in ${\cal N}=4$ SYM),
 which fails to materialize.

There is however still hope for all-loop integrability in other closed {subsectors}. As we have mentioned, one can identify sectors for which the one-loop dilation operator is identical to a restriction
of the ${\cal N}=4$ dilation operator. The largest such sector that remains closed to all orders is the $SU(2,1|2)$ sector,
which consists entirely of letters belonging to the ${\cal N}=2$  vector multiplet,
and it is thus a universal sector present in all ${\cal N}=2$ superconformal gauge theories with a Lagrangian description.
Of course, in any given theory, all the other fields (such as the fundamental hypermultiplets of SCQCD) do
affect the renormalization of the $SU(2,1|2)$ sector, so at sufficiently high order
 the dilation operator {\it will} differ from the one of  ${\cal N}=4$ SYM. Nevertheless,
 consideration of the symmetry structure of the magnon S-matrix and of the holographic sigma model (when available)
 lead us to conjecture in Section 5 that the $SU(2,1|2)$ sector may remain integrable to all orders. The simplest scenario
 is that, in any given theory, the dilation operator in this sector coincides with the one in ${\cal N}=4$ SYM,
 up to a model-dependent redefinition of the 't Hooft coupling \cite{LiendoElli} -- a mild 
 but still non-trivial deformation.
 Analogous (though less compelling) speculations
 apply to the universal $SU(2,1|1)$ sector that is present in any ${\cal N}=1$ superconformal gauge theory,
 and even  to the purely bosonic $SU(2,1)$ sector of QCD, near the Banks-Zaks fixed point at the upper edge of the conformal window.

\section{Preliminaries: symmetry structure of the ${\cal N}=2$ SCQCD spin chain}

The field content of 
${\cal N}=2$ superconformal QCD comprises 
an ${\cal N}=2$ vector multiplet $\{\phi,  \lambda_\alpha^{\ph{\a}\Im}, 
\Fm_{\a \b}\}$ and its conjugate, 
 in the adjoint representation of the $SU(N_c)$ gauge group,
and  $N_f=2N_c$ hypermultiplets $\{Q^\Im, \psi_\a,  \bar{\tilde{\psi}}_{\ad} \, ; {\bar Q}_{\Im},\tilde{\psi}_\a, \bar{\psi}_{\ad}\}$, in the (anti)fundamental representation of $SU(N_c)$.
Here $\alpha = \pm$ and $\dot \alpha = \dot \pm$ are Lorentz indices, and $ \Im = \pm$ an $SU(2)_R$ R-symmetry index. We have suppressed
color and flavor indices.

States of the spin chain are constructed by stringing together color-adjoint  single letters from the vector multiplet, 
 and  color-adjoint two-letter  ``dimers'' from the hypermultiplets, \textit{e.g.} $ \psi_i{\bar Q}^i$,
where $i = 1, \dots N_f$ is a contracted flavor index. Furthermore, each letter can be acted upon by an arbitrary number of covariant derivatives.
 
The ${\cal N}=2$ superconformal group is $SU(2_\alpha,2_{\dot \alpha}|2_\Im)$, where the subscripts 
serve  to emphasize the Lorentz and R-symmetry subgroups: $SU(2_\alpha) \times SU(2_{\dot \alpha}) \times SU(2_\Im) \times U(1)_R \subset SU(2_\alpha,2_{\dot \alpha}|2_\Im)$.
The spin chain vacuum is the chiral state ${\rm Tr }\, \phi^k$. It breaks the superconformal group to the subgroup $PSU(2_{\dot \alpha}|2_{\Im})\times SU(2_\alpha)\ltimes {\mathbb R}$,
 where ${\mathbb R}$ is a central generator that gets identified with the spin chain Hamiltonian. In accordance with Goldstone's theorem, broken symmetry generators are manifested as gapless excitations of the spin chain called magnons. Table \ref{genN2} shows the symmetry generators of the ${\cal N}=2$ superconformal algebra. The diagonal boxed generators correspond to the symmetry preserved by the vacuum while the off-diagonal ones are broken and correspond to Goldstone magnons, which transform in the bifundamental representation of  $PSU(2_{\dot \alpha}|2_{\Im})\times SU(2_\alpha)$. 

\begin{table}
\begin{centering}
\begin{tabular}{c|c|c|c}
\multicolumn{1}{c}{} & \multicolumn{1}{c}{$SU(2_{\bd})$} & \multicolumn{1}{c}{$SU(2_{\Jm})$} & $SU(2_{\b})$\tabularnewline
\cline{2-3}
$SU(2_{\dot{\alpha}})$ & $\dot{\Lm}_{\ad}^{\ph{\a}\, \bd}$ & $\bar{\Qm}_{\Jm\, \ad}$ & $\Dm_{\b \ad}^{\dagger}$\tabularnewline
\cline{2-3}
$SU(2_{\Im})$ & $\bar{\Sm}^{\Im\, \bd}$ & $\Rm_{\Jm}^{\ph{\Jm}\, \Im}$ & $\lambda_{\b}^{\dagger\, \Im}$\tabularnewline
\cline{2-4}
\multicolumn{1}{c}{$SU(2_{\alpha})$} & \multicolumn{1}{c}{$\Dm^{\a \bd}$} & $\lambda_{\Jm}^{\ph{\Jm} \a}$ & \multicolumn{1}{c|}{$\Lm_{\b}^{\ph{\b}\, \a}$}\tabularnewline
\cline{4-4}
\end{tabular}
\par\end{centering}

\caption{\label{genN2}The ${\cal N}=2$ superconformal generators.
 The boxed generators are preserved by the choice of the spin chain vacuum while the unboxed ones  are broken and correspond to  Goldstone excitations. The
broken generators are identified with the corresponding magnon: the
upper-right column contains magnon creation operators while the lower-left row contains magnon annihilation operators.}

\end{table}

A priori, the two-body magnon S-matrix when decomposed according to $SU(2_{\dot \alpha}|2_\Im) \times SU(2_\alpha)$ quantum numbers will take the form
\be
S_{SU(2_{\dot \alpha},2_{\alpha}|2_\Im)}=S_{ SU(2_{\dot \alpha}|2_\Im)   }    \times S^{\bf 1}_{SU(2_{\alpha})}+S^\prime_{SU(2_{\dot \alpha}|2_\Im)}\times S^{\bf 3}_{SU(2_{\alpha})}\, ,
\ee
where the superscripts ${\bf 1}$ and ${\bf 3}$ denote the singlet and triplet $SU(2_\alpha)$ representations. Remarkably, the product
of two fundamental $SU(2|2)$ representations consists of a single irreducible representation,
which implies that the $SU(2|2)$ two-body S-matrix is completely fixed by symmetry, up to an overall phase \cite{Beisert:2005tm}. Thus, the total two-body S-matrix of our model factorizes as
\be
S_{SU(2_{\alpha},2_{\dot \alpha}|2_\Im)}=S_{SU(2_{\dot \alpha}|2_\Im)}\times S_{SU(2_{\alpha})} \,.
\ee
The
$S_{SU(2_{\dot\alpha}|2_\Im)}$ factor is the  two-body S-matrix of the magnons in the $SU(2_\alpha)$ highest weight state, namely  $\{\,\lambda_+^{\ph{+}\Im}, \Dm_{+ \ad}\,\}$,
while $S_{SU(2_\alpha)}$ is the two-body S-matrix of the magnons  in the $SU(2_{\dot\alpha}|2_\Im)$ highest weight state, 
namely $\{\lambda_\alpha^{\ph{\a}+}\}$.

The symmetry analysis also helps us organize the calculation of the dilation generator. We can identify
two ``orthogonal'' all-order closed subsectors,  associated with either factor of the two-body S-matrix.
Exciting an arbitrary number of $SU(2_\alpha)$ highest weight magnons  $\{\,\lambda_+^{\ph{+}\Im}, 
\Dm_{+ \ad}\}$ above the
 spin chain vacuum ${\rm Tr} \, \phi^k$, 
 and demanding closure of the dilation operator,
 we obtain a subsector with enhanced $SU(2,1|2)$ symmetry, spanned by the following letters:
\be \label{SU(2,1|2)sector}
SU(2,1|2)\mbox{ sector:}\qquad\qquad
(\Dm_{+ \ad})^n\{\, \phi, \lambda_+^{\ph{+}\Im}, \Fm_{+ +}\,  \}\,.
\ee
Here the covariant derivatives are understood to be totally symmetrized at each site, so for example 
$ (\Dm_{+ \ad})^n \phi$
is shorthand for ${\cal D}_{+\{ \dot \alpha_1} {\cal D}_{+\dot \alpha_2}\dots {\cal D}_{+\dot \alpha_n\}} \phi$.
The introduction of the self-dual field strength $\Fm_{+ + } = [ {\cal D}_{+ \dot +},  {\cal D}_{+ \dot -}]$
is necessary to achieve closure of the dilation operator because of the transition $ \epsilon_{\Im \Jm} \lambda_+^{\ph{+}\Im} \lambda_+^{\ph{+}\Jm} 
\leftrightarrow \phi \Fm_{+ +}$.

Similarly, considering the $SU(2_{\dot\alpha}|2_\Im)$ highest weight magnons
$\{\lambda_\alpha^{\ph{\a}+}\}$, and demanding closure we obtain a sector with  $SU(2|1)$ symmetry:
\be \label{SU(2|1)sector}
SU(2|1)\mbox{ sector:}\qquad \qquad 
\{\, \phi, \lambda_\alpha^{\ph{\a} +}, {\cal M}^{++}\,\} \, ,
\ee
where we have introduced the notation ${\cal M}^{\Im \Jm} \equiv Q_i^\Im \bar Q^{i \Jm}$. Inclusion of the ${\cal M}^{++}$ dimer
is forced at two loops by the transition $\epsilon^{\a \b}\lambda_\a^{\ph{\a}+} \lambda_\b^{\ph{\b}+} \leftrightarrow \phi  {\cal M}^{++}$.

In the rest of the paper we will consider separately these two subsectors. The $SU(2,1|2)$ sector exists in any ${\cal N}=2$ gauge theory, including
${\cal N}=4$ SYM, while the $SU(2|1)$ sector is special to ${\cal N}=2$ SCQCD and has the potential to reveal a new integrability structure.

\section{The two-loop Hamiltonian in the $SU(2|1)$ sector}
\label{sec:symmetryanalysis}

In this section we will use symmetry arguments to fix the two-loop Hamiltonian of the $SU(2|1)$ sectors, up to a few arbitrary coefficients. With this result at hand, we will proceed in the following section to calculate the two-body scattering of magnons and test integrability of the sector. 
To avoid cluttering we will suppress the ``$+$'' $SU(2)_R$ index
and write the letters as
\be
\label{letters}
\{\, \phi, \lambda_{\a}, \MM\,\}\, .
\ee
At one loop the sector decomposes into $\{\phi, \lambda_\alpha \}$ and $\{\phi, {\cal M} \}$. Each of these subsectors is separately integrable: The first one, because it is identical to the corresponding sector in ${\cal N}=4$ SYM. The second one, because its Hamiltonian turns out to be trivial \cite{Gadde:2010zi}  -- 
the dimer  ${\cal M}$ does not move on the $\phi$ chain so each string of $\phi$'s and ${\cal M}$'s is already an exact eigenstate.
The $SU(2|1)$ sector becomes interesting at two loops, where interaction with $\MM$ affects 
the scattering of the asymptotic $\lambda_\alpha$ magnons.

To avoid an explicit Feynman diagram calculation we will use the approach of \cite{Beisert:2003ys}, where the symmetry algebra was used to restrict the form of the spin chain Hamiltonian in the $SU(2|3)$ subsector of $\NN=4$ SYM. In that case, the two-loop Hamiltonian turned out to be completely fixed by  symmetry.

\subsection*{Parity}

It will be useful to define a ``parity'' operation on the states of the chain. As explained in \cite{Liendo:2011wc}, $\Nm=2$ SCQCD admits a parity transformation that commutes with the Hamiltonian at all loops. The transformations relevant for the fields in the $SU(2|1)$ subsector are
\begin{equation}
 \f^a_{\ph{a} b}  \lra -\f^b_{\ph{b}a}\, ,  \quad  \lambda^a_{\ph{a} b}  \lra -\lambda^b_{\ph{b} a}\, ,
 \quad \MM^a_{\ph{b}b}  \lra -\MM^b_{\ph{b}a} \, .
\end{equation}
This is just transposition of adjoint indices with an extra minus sign. The action on a single trace state is then (using a ket notation for the states of the chain):
\be
P | A_1\, .\, .\, .\, A_L \rangle = (-1)^{L+f(f+1)/2}| A_L\, .\, .\, .\, A_1 \rangle\, ,
\ee
where $f$ is the number of fermionic fields and $L$ is the length of the state considering $\MM$ as a {\it single-site} object.

\subsection{Symmetry analysis}

The states of the sector furnish a representation of the $SU(2|1)$ algebra. In the interacting theory, the symmetry generators can be written as a perturbation series in the coupling constant \cite{Beisert:2003ys,Beisert:2004ry},
\be
\Jm(g)=\sum_{k=0}^{\infty}g^k \Jm_k\, .
\ee
As usual when working with spin chains we will  focus in the \textit{local} action of the generators, the \textit{complete} action  being a sum of local terms. Following \cite{Beisert:2003ys} we will represent the action of a generator by the symbol
\be
\label{Jk}
\Jm_k \sim \PTerm{a_1\ldots a_n}{b_1\ldots b_m}\, .
\ee
This replaces the string of fields $a_1 \ldots a_n$ by $b_1 \ldots b_m$ and gives zero otherwise. To obtain the total action we apply this transformation at each site of the closed chain. For example,
\be
\PTerm{A B}{C D}| A B E A B F \rangle = | C D E A B F \rangle + 0 + 0 + | A B E C D F \rangle + 0 + 0\, .
\ee
Of course, we will pick up an extra minus sign each time a fermionic generator ($\Qm$ or $\Sm$) hops a fermionic field.
An interaction with $n+m$ entries will be said to have $n+m$ legs. Because corrections to the generators have their origin in planar perturbation theory, the number of legs is restricted by the order of the coupling constant we are considering. The counting is easier if we forget for a moment our definition of $\MM$ and consider $Q$ as fundamental field of our sector. The number of legs is then restricted by,
\be
\label{legs}
n+m=k+2\, ,
\ee
where $k$ is the order of the coupling.\footnote{As in \cite{Beisert:2003ys}, we use gauge invariance of cyclic states to increase the legs of the generators to its maximum value, {\it i.e.} $k+2$ at order $k$ in the coupling.} Now, if a $Q$ field sits at the far right in the upper or lower row of \eqref{Jk}, we know that the next field to its right will be a $\bar{Q}$, in order to have a flavor singlet. An analogous analysis holds for a $\bar{Q}$ sitting in the far left. This means that after writing the $\Jm$ generators using the $Q$ and $\bar{Q}$ fields, we can replace all the $Q$'s($\bar{Q}$'s) in the far right(left) with an $\MM$ symbol, in addition to the explicit $Q\bar{Q}=\MM$ replacement.

    \subsection*{The $SU(2|1)$ algebra}

To obtain the $SU(2|1)$ algebra we start from the full $SU(2,2|2)$ generators:\footnote{We follow the conventions of \cite{Liendo:2011xb}.}
\be
\{\, \Lm_{\a}^{\ph{\a}\b}, \dot{\Lm}_{\ad}^{\ph{\ad}\bd}, \Rm_{\Im}^{\ph{\Im}\Jm}, \Pm_{\a\bd}, \Km^{\a\bd}, D, r, \Qm_\a^{\ph{\a}\Im}, \Sm_\Im^{\ph{\Im}\a}, \bar{\Qm}_{\ad\, \Im}, \bar{\Sm}^{\ad\, \Im}\, \}\, ,
\ee
where $\Lm$ and $\dot{\Lm}$ are the Lorentz generators, $\Rm$ and $r$ correspond to $SU(2)_R$ and the $U(1)$ $r$-charge, $D$ is the dilation operator and $\Qm$ and $\Sm$ are the supercharges. We now define
\begin{align}
\Qm_\a & \equiv \Qm_\a^{\ph{\a}+}\, ,
\\
\Sm^\a & \equiv \Sm_+^{\ph{+}\a}\, ,
\\
\Um & \equiv \Rm_{+}^{\ph{+}+} + \sfrac{1}{2}\left(D_0 - r\right)\, ,
\\
\label{dilation}
\delta \Hm & \equiv \delta D\, .
\end{align}
We have split the interacting dilation generator as
\be
D =D_0 + \delta D\, ,
\ee
where $D_0$ measures the classical conformal dimension and $\delta D$ its quantum corrections.\footnote{To be consistent with (\ref{dilation}) we also define $\Hm_0 \equiv D_0$, although $\Hm_0$ is not an $SU(2|1)$ generator.}
The $SU(2|1)$ generators are then:
\be
\Jm=\{\Lm_{\a}^{\ph{\a}\b}, \Um,  \delta \Hm, \Qm_\a, \Sm^\a\}\, .
\ee
As in \cite{Beisert:2003ys}, we enhanced the algebra by the extra central $U(1)$ generator $\delta H$. The commutation relations are easy to obtain from the original $SU(2,2|2)$ commutators. Generators carrying $SU(2)$ Lorentz indices transform canonically according to:
\begin{align}
[\Lm_{\a}^{\ph{\a}\b},\Jm_\g] & = \delta^\b_\g \Jm_\a -\sfrac{1}{2}\delta^\b_\a \Jm_\g\,,
& [\Lm_{\a}^{\ph{\a}\b},\Jm^\g] & = -\delta^\g_\a \Jm_\b +\sfrac{1}{2}\delta^\b_\a \Jm^\g\, .
\end{align}
The only non-zero anti-commutator is:
\be
\{\Sm^\b,\Qm_\a\} = \Lm_\a^{\ph{\a}\b}+\delta^\b_\a(\Um +\sfrac{1}{2}\delta \Hm)\, 
\ee
and the non-zero $\Um$-charges are:
\begin{align}
[\Um, \Qm_\a] & = -\sfrac{1}{2}\Qm_\a \, , & [\Um, \Sm^\a] & = \sfrac{1}{2}\Sm^\a\, .
\end{align}
Also,
\be
[\Jm, \delta \Hm] = 0\, ,
\ee
confirming that $\delta \Hm$ is indeed a central element. 

Note that $\Um$ is defined in terms of generators that do not receive quantum corrections and therefore it will not be modified in the interacting theory. The same applies to $\Lm_{\a}^{\ph{\a}\b}$ if we choose a regularization scheme consistent with Lorentz symmetry. In general, different regularization schemes
can differ in which generators will be quantum deformed, but the physical outcome (in this case, the eigenvalues of the dilation operator)
must of course be the same in all schemes. Our algebraic analysis takes the simplest form in a scheme where the Lorentz generators maintain the tree level
form. An example of such a scheme is dimensional regularization, where Lorentz invariance is manifest at each step.

\subsection{The interacting generators}

The tree-level representation of the $SU(2|1)$ algebra reads
\bea
\Um\eq\PTerm{\f}{\f}+\sfrac{1}{2}\PTerm{\a}{\a}\, ,
\nln
\Lm_{\alpha}^{\ph{\a}\beta}\eq \PTerm{\alpha}{\beta}-\sfrac{1}{2}\delta^\alpha_\beta\PTerm{\gamma}{\gamma}\, ,
\nln
(\Qm_{\alpha})_0\eq e^{i\beta_1}\PTerm{\f}{\alpha}\, ,
\nln
(\Sm^{\alpha})_0\eq e^{-i\beta_1}\PTerm{\a}{\f}\, ,
\eea
where the subscript ``0'' indicates that we are working at tree level. The idea is to consider perturbative deformations of these generators and restrict their form using the $SU(2|1)$ algebra. 
In principle, there should be fluctuations in the length, but because we consider the dimeric impurity $\MM$ as a single-site object, the length always stays constant.
For $\Hm_2$ we have:
\be
\begin{split}
\Hm_2 & =
c_0\PTerm{\f \f}{\f \f}+ c_1\PTerm{\f \MM}{\f \MM}+c_2\PTerm{\MM \f}{\MM \f}
+c_3\PTerm{\MM}{\MM}
+c_4\PTerm{\f \a}{\f \a}+c_5\PTerm{\a \f}{\a \f}
\\
& \quad +c_6\PTerm{\f \a}{\a \f}+c_7\PTerm{\a \f}{\f \a}+c_8\PTerm{\a \MM}{\a \MM}+c_9\PTerm{\MM \a}{\MM \a}+c_{10}\PTerm{\a \beta}{\a \beta}+c_{11}\PTerm{\a \beta}{\beta \a}\, .
\end{split}
\ee
Imposing invariance under parity we obtain:
\be
c_1=c_2\, , \qquad c_4=c_5\, , \qquad c_6=c_7\, , \qquad c_8=c_9\, .
\ee
In addition, protection of $\phi \phi$ implies $c_0=0$.\footnote{In \cite{Beisert:2003ys} this condition was obtained using the algebra constraints, in our case we have to give it as extra input.} This still leaves seven independent coefficients. Imposing that the algebra commutation relations are satisfied perturbatively eliminates six of them, leaving us with one undetermined parameter, $c_1 \equiv \a_1^2$, which is associated with a rescaling of the coupling and cannot be fixed by algebraic means. The procedure is now completely algorithmic and it was described in detail in \cite{Beisert:2003ys}. For each perturbative correction we consider the most general ansatz consistent with conservation of classical energy, $r$-charge and equation \eqref{legs}. Consistency of the algebra commutations relations significantly reduces the number of independent parameters. As extra input we use the fact that in the $SU(1|1)$ subsector spanned by $\{\, \phi,\lambda_+\,\}$ the two-loop Hamiltonian of $\Nm=2$ SCQCD should be identical to the corresponding Hamiltonian in $\Nm=4$ SYM \cite{Pomoni:2011}.
We present our results in Tables \ref{quanham} and \ref{fermgen}. At first sight, there seems to be a high number of independent coefficients, however most of them are unphysical.
The two coefficients $\{\,\a_1,\a_3\,\}$ can be reabsorbed by a redefinition of the coupling,\footnote{Note of course that $\alpha_1 \neq 0$, otherwise the whole one-loop Hamiltonian ${\cal H}_2$ would vanish. The actual value of $\alpha_1$ could be fixed by comparison with the explicit perturbative calculation \cite{Liendo:2011xb}:  ${\cal H}_{here} = D_{there}$, and $\alpha_1^2 = 2$.
}
\be
g\rightarrow \a_1 g+\a_3 g^3\, .
\ee
The six coefficients $\{\,\beta_1, \beta_2, \delta_1, \delta_2, \delta_3, \delta_4\,\}$ correspond to similarity transformations and never show up in physical quantities like anomalous dimensions or S-matrix elements. We are then left with $\{\,\eta$, $\chi\,\}$ which do show up in physical quantities and therefore cannot be ignored. However, the S-matrix elements that we will study in the next section happen to be independent of $\{\,\eta$, $\chi\,\}$.

\begin{table}
\bea
\Hm_0\eq
\PTerm{\f}{\f}+\small{2}\PTerm{\MM}{\MM}+\sfrac{3}{2}\PTerm{\a}{\a},
\nln
\Hm_2\eq
 \alpha_1^2\bigbrk{\PTerm{\f \MM}{\f \MM}+\PTerm{\MM \f}{\MM \f}}
+2\alpha_1^2\PTerm{\MM}{\MM}
+\alpha_1^2\bigbrk{\PTerm{\f \a}{\f \a}+\PTerm{\a \f}{\a \f}}
-\alpha_1^2\bigbrk{\PTerm{\f \a}{\a \f}+\PTerm{\a \f}{\f \a}}
\nl
+\alpha_1^2\bigbrk{\PTerm{\a \MM}{\a \MM}+\PTerm{\MM \a}{\MM \a}}+\alpha_1^2\PTerm{\a \beta}{\a \beta}+\alpha_1^2\PTerm{\a \beta}{\beta \a},
\nln
\Hm_3\eq
-\alpha_1^3\,e^{i\beta_2}\,\varepsilon_{\alpha\beta}\bigbrk{\PTerm{\alpha\beta}{\f \MM}+\PTerm{\alpha\beta}{\MM \f}}
-\alpha_1^3\,e^{-i\beta_2}\,\varepsilon^{\alpha\beta}\bigbrk{\PTerm{\f \MM}{\alpha\beta}+\PTerm{\MM \f}{\alpha\beta}},
\nln
\Hm_4\eq
(-\sfrac{3}{2}\a_1^4+2\a_1\a_3)\bigbrk{\PTerm{\f \f \a}{\f \f \a}+\PTerm{\a \f \f}{\a \f \f}}
+(\a_1^4-\a_1\a_3)\bigbrk{\PTerm{\f \f \a}{\f \a \f}+\PTerm{\a \f \f}{\f \a \f}}
\nl
-\sfrac{1}{2}\a_1^2\bigbrk{\PTerm{\f \f \a}{\a \f \f}+\PTerm{\a \f \f}{\f \f \a}}
+(\a_1^4-\a_1\a_3)\bigbrk{\PTerm{\f \a \f}{\a \f \f}+\PTerm{\f \a \f}{\f \f \a}}
\nl
+(-\sfrac{5}{4}\a_1^2+\a_1\a_3-\eta+\chi)\bigbrk{\PTerm{\f \f \MM}{\f \f \MM}+\PTerm{\MM \f \f}{\MM \f \f}}
\nl
+(-\sfrac{31}{4}\a_1^2+7\a_1\a_3+\chi)\bigbrk{\PTerm{\f \MM}{\f \MM}+\PTerm{\MM \f}{\MM \f}}
+(\a_1^4-2\a_1\a_3+\eta)\bigbrk{\PTerm{\f \MM}{\MM \f}+\PTerm{\MM \f}{\f \MM}}
\nl
+(\sfrac{19}{2}\a_1^4-10\a_1\a_3+2\eta-2\chi)\PTerm{\MM \f \MM}{\MM \f \MM}+2\eta\PTerm{\MM \MM}{\MM \MM}
\nl
+(-2\a_1^4+2\a_1\a_3-\eta+\chi+i\a_1^2(\delta_1+\delta_2))\bigbrk{\PTerm{\a \f \MM}{\f \a \MM}+\PTerm{\MM \f \a}{\MM \a \f}}
\nl
+(-2\a_1^4+2\a_1\a_3-\eta+\chi-i\a_1^2(\delta_1+\delta_2))\bigbrk{\PTerm{\f \a \MM}{\a \f \MM}+\PTerm{\MM \a \f}{\MM \f \a}}
\nl
+(-\sfrac{13}{4}\a_1^4+3\a_1\a_3-\eta+\chi)\bigbrk{\PTerm{\f \a \MM}{\f \a \MM}+\PTerm{\MM \a \f}{\MM \a \f}}
\nl
+(-2\a_1^4+2\a_1\a_3+\eta)\bigbrk{\PTerm{\a \MM}{\a \MM}+\PTerm{\MM \a}{\MM \a}}
+(2\a_1^4-2\a_1\a_3+\eta)\bigbrk{\PTerm{\a \MM}{\MM \a}+\PTerm{\MM \a}{\a \MM}}
\nl
+(-\sfrac{1}{4}\a_1^4+\a_1\a_3)\bigbrk{\PTerm{\f \a \gb}{\f \a \gb}+\PTerm{\gb \a \f}{\gb \a \f}}
+(-\sfrac{7}{4}\a_1^4+\a_1\a_3)\bigbrk{\PTerm{\f \a \gb}{\f \gb \a}+\PTerm{\gb \a \f}{\a \gb \f}}
\nl
+(\a_1^4-\a_1\a_3-i\a_1^2\delta_1)\bigbrk{\PTerm{\f \a \gb}{\a \f \gb}+\PTerm{\gb \a \f}{\gb \f \a}}
+(\a_1^4-\a_1\a_3+i\a_1^2\delta_1)\bigbrk{\PTerm{\a \f \gb}{\f \a \gb}+\PTerm{\gb \f \a}{\gb \a \f}}
\nl
+(\sfrac{1}{4}\a_1^4+i\a_1^2\delta_3)\bigbrk{\PTerm{\f \a \gb}{\gb \f \a}+\PTerm{\gb \a \f}{\a \f \gb}}
+(\sfrac{1}{4}\a_1^4-i\a_1^2\delta_3)\bigbrk{\PTerm{\gb \f \a}{\f \a \gb}+\PTerm{\a \f \gb}{\gb \a \f}}
\nl
+(-\sfrac{7}{2}\a_1^4+4\a_1\a_3)\PTerm{\a \f \gb}{\a \f \gb}+\sfrac{1}{2}\a_1^2\PTerm{\a \f \gb}{\gb \f \a}
\nl
+(-\sfrac{7}{2}\a_1^4+4\a_1\a_3-\eta+\chi)\bigbrk{\PTerm{\MM \a \gb}{\MM \a \gb}+\PTerm{\gb \a \MM}{\gb \a \MM}}
\nl
+(\sfrac{3}{2}\a_1^4-2\a_1\a_3+\eta-\chi)\bigbrk{\PTerm{\MM \a \gb}{\MM \gb \a}+\PTerm{\gb \a \MM}{\a \gb \MM}}
\nl
+(-\sfrac{9}{4}\a_1^4+3\a_1\a_3)\bigbrk{\PTerm{\a \gb \g}{\a \g \gb}+\PTerm{\g \gb \a}{\gb \g \a}}
+(\sfrac{1}{2}\a_1^4-2\a_1\a_3)\bigbrk{\PTerm{\a \gb \g}{\gb \g \a}+\PTerm{\g \gb \a}{\a \g \gb}}
\nl
+(-\sfrac{1}{2}\a_1^4+2\a_1\a_3)\PTerm{\a \gb \g}{\g \gb \a}\, .
\nn
\eea
\caption{The Hamiltonian up to order $g^4$.}
\label{quanham}
\end{table}

\begin{table}
\bea
(\Qm_{\alpha})_0\eq e^{i\beta_1}\PTerm{\f}{\alpha}\, ,
\nln
(\Qm_{\alpha})_1\eq \,\alpha_1\,e^{i(\beta_1+\beta_2)}\varepsilon_{\alpha\beta}\PTerm{\beta}{\MM},
\nln
(\Qm_{\alpha})_2\eq
i e^{i\beta_1}(\delta_1+\delta_2+\delta_4)\bigbrk{\PTerm{\f \f}{\f \a}+\PTerm{\f \f}{\a \f}}
+ e^{i\beta_1}(\sfrac{1}{4}\a_1^2+i \delta_4)\bigbrk{\PTerm{\f \MM}{\a \MM}+\PTerm{\MM \f}{\MM \a}}
\nl
+ e^{i\beta_1}(\sfrac{1}{4}\a_1^2+i \delta_3)\bigbrk{\PTerm{\f \gb}{\gb \a}-\PTerm{\gb \f}{\a \gb}}
+ i e^{i\beta_1}(\delta_2 + \delta_4)\bigbrk{\PTerm{\f \gb}{\a \gb}-\PTerm{\gb \f}{\gb \a}},
\nln
(\Sm^{\alpha})_0\eq e^{-i\beta_1}\PTerm{\a}{\f}\, ,
\nln
(\Sm^{\alpha})_1\eq \,\alpha_1\,e^{-i(\beta_1+\beta_2)}\varepsilon^{\alpha\beta}\PTerm{\MM}{\beta},
\nln
(\Sm^{\alpha})_2\eq
-i e^{-i\beta_1}(\delta_1+\delta_2+\delta_4)\bigbrk{\PTerm{\f \a}{\f \f}+\PTerm{\a \f}{\f \f}}
+ e^{-i\beta_1}(\sfrac{1}{4}\a_1^2-i \delta_4)\bigbrk{\PTerm{\a \MM}{\f \MM}+\PTerm{\MM \a}{\MM \f}}
\nl
+ e^{-i\beta_1}(\sfrac{1}{4}\a_1^2-i \delta_3)\bigbrk{\PTerm{\gb \a}{\f \gb}-\PTerm{\a \gb}{\gb \f}}
- i e^{-i\beta_1}(\delta_2 + \delta_4)\bigbrk{\PTerm{\a \gb}{\f \gb}-\PTerm{\gb \a}{\gb \f}}.
\nn
\eea
\caption{Fermionic $SU(2|1)$ generators up to order $g^2$.}
\label{fermgen}
\end{table}

\section{The magnon S-matrix in the $SU(2|1)$ sector }
\label{sec:dispersionRelation}

We now proceed to  calculate the magnon two-body S-matrix in the $SU(2|1)$ sector, and to check whether it satisfies the Yang-Baxter equation.
Let us start by defining the momentum eigenstate of a single  excitation,
  \begin{equation}
    \ket{\lambda_\alpha(p)}=\sum_k e^{ipk}\ket{\alpha_k}\, ,
  \end{equation}
  where $k$ labels the position of the particle,
  \begin{equation}
    \ket{\alpha_k}=\ket{\ldots\phi\MMM{\lambda_\alpha}{k}\phi\ldots}\,.
  \end{equation}
Its dispersion relation is easily obtained by acting with the Hamiltonian:
  \begin{equation}
    \Hm\ket{\lambda_\alpha(p)}=g^2\alpha^2_1\left[(2-e^{ip}-e^{-ip})
                             +g^2\alpha^2_1(-3+2(e^{ip}+e^{-ip})-\half(e^{2ip}+e^{-2ip}))\right]\ket{\lambda_\alpha(p)}\, ,
  \end{equation}
  hence,
  \begin{equation}
    \begin{split}
      E^\lambda(p)&=4(g^2\alpha^2_1-2g^4\alpha^4_1)\sin^2\frac{p}{2}+2g^4\alpha^4_1\sin^2p + O(g^6)\, .
    \end{split}
  \end{equation}
 To extract the S-matrix we will use the familiar perturbative asymptotic Bethe ansatz, see {\it e.g.}  \cite{Staudacher:2004tk}. For the $SU(2_\alpha)$ singlet two-body state we define:
  \begin{equation}
  \label{singletket}
    \begin{split}
      \ket{\lambda_{[\alpha}\lambda_{\beta]}}
      =&\sum_{k<l-1}\SPsi(k,l)\ket{\ldots\phi\MMM{\lambda_{[\alpha}}{k}\phi\ldots\phi\MMM{\lambda_{\beta]}}{l}\phi\ldots}\\
       &+\sum_k\nPsi(k)\ket{\ldots\phi\MMM{\lambda_{[\alpha}}{k}\MMM{\lambda_{\beta]}}{k+1}\phi\ldots}
       +\sum_k\MPsi(k)\ket{\ldots\phi\MMM{\MM}{k}\phi\ldots}\, ,
    \end{split}
  \end{equation}
valid up to order $g^2$. The $\Psi$'s correspond Schr\"{o}dinger wave functions and $k$ and $l$ label the positions of the particles in the $\phi$ vacuum. 
At this order in perturbation theory a transition $\lambda_{[\alpha}\lambda_{\beta]} \rightarrow \MM$ is possible and this is taken into account by the last term in \eqref{singletket}. In order to solve the scattering problem we consider the following ansatz:
  \begin{equation}
    \label{eq:DefOfPsi}
    \begin{split}
      \SPsi(k,l)&=e^{i(p_1k+p_2l)}+\SS(p_2,p_1)e^{i(p_1l+p_2k)}\, ,
      \\
      \nPsi(k)&=\nS(p_2,p_1)e^{i(p_1+p_2)k}\, ,
      \\
      \MPsi(k)&=\MS(p_2,p_1)e^{i(p_1+p_2)k}\, .
    \end{split}
  \end{equation}
Here $\SS(p_2,p_1)$, $\nS(p_2,p_1)$ and $\MS(p_2,p_1)$ are functions of $g$ and represent the different scattering amplitudes. Imposing the Schr\"{o}dinger equation
  \begin{equation}
    \Hm\ket{\lambda_{[\alpha}\lambda_{\beta]}}=E(p_1,p_2)\ket{\lambda_{[\alpha}\lambda_{\beta]}}\, ,
  \end{equation}
  for the separate cases $l > k+2$, $l=k+2$ and $l=k+1$ we can solve
 for the scattering amplitudes to order  $g^2$. The interesting term is $\SS(p_2,p_1)$, which governs
 the asymptotic magnon scattering,
  \begin{equation}
    \begin{split}
      \SS(p_2,p_1)=&-\frac{1-2e^{ip_2}+e^{i(p_1+p_2)}}{1-2e^{ip_1}+e^{i(p_1+p_2)}}\\
      \times&\Big(1+2ig^2\alpha^2_1
      \frac{(\cos p_1-2\cos(p_1-p_2)+\cos p_2)\sin{\frac{p_1}{2}}\sin{\frac{p_2}{2}}(\sin p_1-\sin p_2)}
      {\cos(\frac{p_1-p_2}{2})(3-2\cos p_1-2\cos p_2+\cos(p_1+p_2))} + O(g^4)\Big) .
    \end{split}
  \end{equation}
In 
the triplet sector the ansatz is simpler
 since $\lambda_{\{\alpha}\lambda_{\beta\}}$ does not mix with $\mathcal{M}$,
 \begin{equation}
    \begin{split}
      \ket{\lambda_{\{\alpha}\lambda_{\beta\}}}
      =&\sum_{k<l-1}\TPsi(k,l)\ket{\ldots\phi\MMM{\lambda_{\{\alpha}}{k}\phi\ldots\phi\MMM{\lambda_{\beta\}}}{l}\phi\ldots}
        +\sum_{k}\Psi_{\rep{3}n}(k)\ket{\ldots\phi\MMM{\lambda_{\{\alpha}}{k}\MMM{\lambda_{\beta\}}}{k+1}\phi\ldots}  \, ,
    \end{split}
  \end{equation}
  where
  \begin{equation}
    \label{eq:DefOfPsi}
    \begin{split}
      \TPsi(k,l)&=e^{i(p_1k+p_2l)}+\TS(p_2,p_1)e^{i(p_1l+p_2k)}\, ,
      \\
      \Psi_{\rep{3}n}(k)&=S_{\rep{3}n}(p_2,p_1)e^{i(p_1+p_2)k}\,.
    \end{split}
  \end{equation}
 We find
  \begin{equation}
    \begin{split}
      \TS(p_2,p_1)=&-1
      -ig^2\alpha^2_1(\sin{p_1}-\sin{(p_1-p_2)}-\sin{p_2}) + O(g^4)\,.
    \end{split}
  \end{equation}

\subsection*{Checking the Yang-Baxter equation}

We are finally ready to check the Yang-Baxter equation
 for the  two-body magnon S-matrix. The  equation reads (see Figure 1 for the index flow)
\begin{equation}
  S^{\delta\epsilon}_{\alpha\beta}(p_1,p_2)S^{\tau\gamma'}_{\epsilon\gamma}(p_1,p_3)S^{\alpha'\beta'}_{\delta\tau}(p_2,p_3)
  =S^{\beta'\gamma'}_{\epsilon\delta}(p_1,p_2)S^{\alpha'\epsilon}_{\alpha\tau}(p_1,p_3)S^{\tau\delta}_{\beta\gamma}(p_2,p_3) \,.
\end{equation}
\bigskip
    \begin{figure}[h]
             \begin{center}
            \includegraphics[scale=0.8]{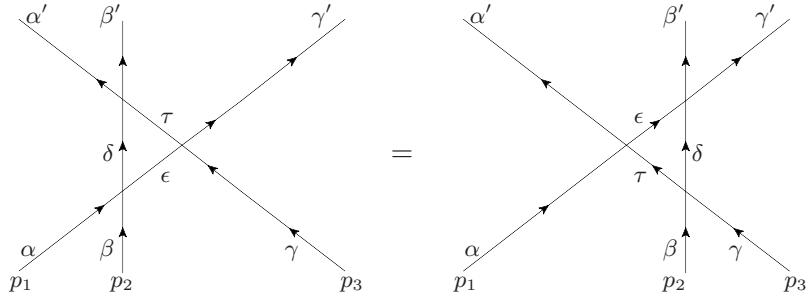}
            \put(-292,7){{\footnotesize $\alpha$}}
            \put(-296,-5){{\footnotesize $p_1$}}
            \put(-262,7){{\footnotesize $\beta$}}
            \put(-258,-5){{\footnotesize $p_2$}}
            \put(-192,7){{\footnotesize $\gamma$}}
            \put(-171,-5){{\footnotesize $p_3$}}
            \put(-239,35){{\footnotesize $\epsilon$}}
            \put(-239,57){{\footnotesize $\tau$}}
            \put(-261,43){{\footnotesize $\delta$}}
            \put(-290,95){{\footnotesize $\alpha'$}}
            \put(-262,95){{\footnotesize $\beta'$}}
            \put(-182,95){{\footnotesize $\gamma'$}}  
            \put(-152,43){=}
            \put(-124,7){{\footnotesize $\alpha$}}
            \put(-128,-5){{\footnotesize $p_1$}}
            \put(-49,7){{\footnotesize $\beta$}}
            \put(-49,-5){{\footnotesize $p_2$}}
            \put(-24,7){{\footnotesize $\gamma$}}
            \put(-3,-5){{\footnotesize $p_3$}}
            \put(-60,35){{\footnotesize $\tau$}}
            \put(-60,57){{\footnotesize $\epsilon$}}
            \put(-38,43){{\footnotesize $\delta$}}
            \put(-122,95){{\footnotesize $\alpha'$}}
            \put(-49,95){{\footnotesize $\beta'$}}
            \put(-14,95){{\footnotesize $\gamma'$}}    
            \vspace{0.5cm}
              \caption{Diagrammatic representation of the Yang-Baxter equation.}
              \label{YB}
            \end{center}
            \end{figure}

\noindent               
Defining:
\bea
A(p_1,p_2) & = &  \TS(p_1,p_2)\, ,
\\
B(p_1,p_2) & = &  \frac{1}{2}(\SS(p_1,p_2)- \TS(p_1,p_2))\, ,
\eea
we can rewrite the S-matrix in terms of the identity operator $\mathbb{I}$ and the
trace operator $\mathbb{K}$,
\begin{equation}
  S(p_1,p_2)=A(p_1,p_2)\mathbb{I}+B(p_1,p_2)\mathbb{K}\, .
\end{equation}
As explained {\it e.g.} in \cite{Gadde:2010zi}, the Yang-Baxter equation is equivalent to the single constraint 
\bea
\nn
0 & \stackrel{?}{=} & 2 B(p_1, p_2) A(p_1, p_3) B(p_2, p_3) +
 A(p_1, p_2) A(p_1, p_3) B(p_2, p_3) + B(p_1, p_2) A(p_1, p_3) A(p_2, p_3)
 \\
 & &
+  B(p_1, p_2) B(p_1, p_3) B(p_2, p_3) - A(p_1, p_2) B(p_1, p_3) A(p_2, p_3)\, .
\eea
A necessary condition for  factorization of many-body scattering is the vanishing of the right-hand side. However, working at order $g^2$ we obtain
\be
64 i \a_1^2 e^{i(p_1+p_2+p_3)}\frac{\sin{(\frac{p_1}{2})}^2\sin{(\frac{p_2}{2})}^2\sin{(\frac{p_3}{2})}^2\tan{(\frac{p_1-p_2}{2})}\tan{(\frac{p_1-p_3}{2})}\tan{(\frac{p_2-p_3}{2})}}{(1+e^{i(p_1+p_2)}-2e^{i p_2})(1+e^{i(p_1+p_3)}-2e^{i p_3})(1+e^{i(p_2+p_3)}-2e^{i p_3})}\, ,
\ee
which is certainly non-zero.\footnote{The only solution is the trivial solution $\alpha_1 \equiv 0$, which sets to zero the whole interacting Hamiltonian,
see Table 2.} Failure of the Yang-Baxter equation conclusively shows that the $SU(2|1)$ sector is not integrable at two loops.

.

\section{The universal $SU(2,1|2)$ sector}

The $SU(2,1|2)$ sector (\ref{SU(2,1|2)sector}) consists entirely of letters that belong to the ${\cal N}=2$ vector multiplet,
and it is then present in any ${\cal N}=2$ gauge theory.  Diagrammatic arguments~\cite{Pomoni:2011}  show that
the planar dilation operator in this sector is the same up to {\it two loops} in any ${\cal N}=2$ superconformal theory,
as it  coincides to that order with a restriction of the ${\cal N}=4$ SYM dilation operator. The model
dependence kicks in at three loops.\footnote{In the context of ${\cal N}=4$ SYM, the  $SU(2,1|2)$ sector can be regarded as a non-compact cousin of the $SU(2|3)$ sector, whose  Hamiltonian 
was determined up to three loops by Beisert \cite{Beisert:2003ys} using symmetry arguments. The Hamiltonian of non-compact sectors is much harder to fix. 
Zwiebel's paper \cite{Zwiebel:2005er}
represents the state of the art. 
} 

Choosing the usual chiral vacuum
${\rm Tr} \,  \phi^k$, the Goldstone magnons $\{\,\lambda_+^{\ph{+}\Im\,}, 
\Dm_{+ \ad}\}$
transform in the fundamental representation of $SU(2_{\dot \alpha}| 2_{\Im})$.
Their two-body S-matrix $S_{SU (2_{\dot \alpha}| 2_{\Im})}$  is uniquely determined up to an overall phase by the $SU(2|2)$ symmetry \cite{Beisert:2005tm}, and
thus, just as is the case in ${\cal N}=4$ SYM, it automatically satisfies the Yang-Baxter equation.
This is a first hint 
to suspect that this sector may be generically integrable, at least in the sense of the asymptotic
Bethe ansatz on the infinite chain.\footnote{We are
postponing at this stage the harder questions about finite-size effects.
} 
Of course, factorization of the $n$-body S-matrix into two-body S-matrices is a stronger condition than Yang-Baxter,
and an explicit test at three loops will be required.  A three-loop diagrammatic analysis is in progress \cite{LiendoElli}. 
The strongest conjecture  \cite{LiendoElli} suggested by this perturbative study is that the $SU(2,1|2)$ Hamiltonian of any ${\cal N}=2$ superconformal gauge theory can be
mapped to that of ${\cal N}=4$ SYM by a redefinition of the 't Hooft coupling, 
$g^2 \to f(g^2) = g^2 + O(g^6)$. This would be a trivial operation from the viewpoint of the integrable structure.
Indeed recall that it is still somewhat of a mystery  why the dispersion relation of the ${\cal N}=4$ SYM magnons
takes the  exact form
\be
\Delta - |r| = \sqrt{1 + 8 g^2 \sin^2 \frac{p}{2}}\, ,
\ee
while integrability alone would be compatible with the replacement $g^2 \to f(g^2)$ (which is indeed what happens in the ABJM model \cite{Aharony:2008ug}).
However a redefinition of $g$ can have drastic dynamical consequences, for example it may radically change
the strong coupling behavior of anomalous dimensions (ABJM is again a case in point.)

A second indication in favor
of integrability of the $SU(2,1|2)$ sector comes from the AdS/CFT correspondence -- at least, that is, for the subset of models that admit a string dual.
The simplest ${\cal N}=2$ theories with a known
 string description are the orbifolds of ${\cal N}=4$ SYM by a discrete subgroup {$\Gamma \subset SU(2) \subset SU(4)_R$},
 which are dual to the IIB backgrounds  $AdS_5 \times S^5/\Gamma$  \cite{Kachru:1998ys, Lawrence:1998ja}.
These are quiver gauge theories with product gauge group $SU(N)^k$, where $k$ is the order of $\Gamma$.
The $k$ gauge couplings  are
exactly marginal parameters. If all  gauge couplings  are equal, the spin chain (and the dual sigma model) is completely integrable \cite{Beisert:2005he, Solovyov:2007pw},
 but when they are  different,  integrability of the full chain
  is  broken.\footnote{
   For the simplest example of the $\mathbb{Z}_2$ orbifold,  this phenomenon was studied in detail in  \cite{Gadde:2010zi,Pomoni:2011,Gadde:2010ku},
   which focussed 
 on the magnons transforming in the bifundamental representation of the $SU(N_{c}) \times SU(N_{\check c})$ gauge group, with $N_c \equiv N_{\check c}$.
 For $\lambda \neq \check \lambda$
  their dispersion relation develops a gap.
 The form of their two-body S-matrix is fixed
 by symmetry, and fails to satisfy the Yang-Baxter equation except when $\lambda = \check \lambda$.}
  However, the situation is much better 
     in the $SU(2,1|2)$ sector.\footnote{There are actually $k$ separate $SU(2,1|2)$ sectors, one for each of the $SU(N)$ vector multiplets.}
  At  strong coupling one can study the
  S-matrix of the $SU(2|2)$ excitations using the dual sigma model. 
     Changing the relative gauge couplings is dual  to  twisted-sector deformations in the sigma model:
  to leading order in $\alpha'$ (tree level in the sigma model) they do not change
  the scattering of the $SU(2|2)$ excitations, 
  which live in directions of the target space unaffected by the orbifold. So the $n$-body S-matrix still factorizes
  into two-body S-matrices.
  To be more precise, the only effect  of the twisted deformation
 felt by the $SU(2|2)$ excitations is a renormalization of the string tension. For example, in the $\mathbb{Z}_2$ case,
 the relation between $\alpha'$ and the AdS radius $R$ reads
 \be \label{tensionrenorm}
 \frac{R^4}{\alpha'} = \frac{2 \lambda \check \lambda}{ \lambda + \check \lambda} \, ,
 \ee
where $\lambda$ and $\check \lambda$ are the two 't Hooft couplings.
It would be very interesting to confirm this picture to next order in $\alpha'$, 
where the effect of the twisted deformation is non-trivial, by an explicit one-loop calculation
of the sigma-model S-matrix. Recall that the two-body $SU(2|2)$ S-matrix is completely fixed by symmetry,
so to really probe integrability one would have to study factorization of the $n$-body S-matrix or devise some other test.

In summary, the $SU(2,1|2)$ sector(s) of ${\cal N}=2$ superconformal gauge theories have
the same Hamiltonian as in ${\cal N}=4$ SYM for small $\lambda$ (to two-loop order, $O(\lambda^2)$);
and in theories with AdS duals, the large $\lambda$ limit of the Hamiltonian is also the same as in ${\cal N}=4$ SYM,
modulo a renormalization of the coupling. For example, in the $\mathbb{Z}_2$ quiver theory, it follows from (\ref{tensionrenorm}) that  for large $\lambda$ and large $\check \lambda$
(with $\lambda / \check \lambda$ fixed) the dilation operator in the $SU(2,1|2)$ sector coincides with the one in ${\cal N}=4$ SYM
if one replaces $\lambda \to  2 \lambda \check \lambda /( \lambda + \check \lambda)$.\footnote{This correspondence  is also precisely confirmed \cite{Wenbin}
by considering the strong coupling limit of the matrix model \cite{Pestun:2007rz} that calculates the expectation value of the
1/2 BPS circular Wilson loop in the $\mathbb{Z}_2$ quiver theory, following \cite{Rey:2010ry, Passerini:2011fe}.}
We are led to conjecture
that this remains true for all intermediate values of the coupling, with the appropriate redefinition $\lambda \to f(\lambda)$
that matches the weak and strong coupling behaviors.

\subsection*{$SU(2,1|1)$ and $SU(2,1)$}

In closing, it is tempting to entertain the natural extrapolations of this conjecture to ${\cal N}=1$ and ${\cal N}=0$ conformal gauge theories.
Every ${\cal N}=1$ superconformal gauge theory contains a closed $SU(2,1|1)$ sector, with letters belonging entirely to the ${\cal N}=1$ vector
multiplet,
\be \label{SU(2,1|1)sector}
SU(2,1|1)\mbox{ sector:}\qquad\qquad
(\Dm_{+ \ad})^n\{\, \lambda_+, \Fm_{+ +}\,  \}\,.
\ee
The diagrammatic arguments of \cite{Pomoni:2011}  show again that in any ${\cal N}=1$ superconformal theory the dilation operator in this sector coincides
up to two loops with  the restriction of the ${\cal N}=4$ SYM dilation operator. (Of course this is a meaningful statement only for ${\cal N}=1$ SCFTs that have a weak coupling limit). Choosing the chiral vacuum ${\rm Tr}\, \lambda_{+}^k$, the asymptotic excitations on the chain
are the massless magnons $\{ \Dm_{+ \ad}  \}$, transforming as a doublet of $SU(2_{\dot \alpha})$. This is not enough symmetry
to completely fix the form of the two-body magnon S-matrix, which makes  integrability of the $SU(2,1|1)$ sector somewhat less compelling as a general conjecture.
For  models that admit string duals, some evidence for integrability comes again from the AdS/CFT correspondence.
For example, while the generic Leigh-Strassler deformation of ${\cal N}=4$ SYM is not fully integrable (see \cite{Zoubos:2010kh} for a review),
there is still hope for integrability in the $SU(2,1|1)$ sector. Indeed, 
one can  argue for integrability at strong coupling (to leading order):
the deformation of the $AdS_5 \times S^5$ background that corresponds to the Leigh-Strassler deformation (whatever its explicit form may be)
is not expected to affect the tree-level scattering of excitations in  
 the $SU(2,1|1)$ subsector, since those excitations live entirely in $AdS_5$.

It would be particularly interesting to explore this conjecture in ${\cal N}=1$ super QCD, in the conformal window $\frac{3}{2} N_c < N_f < 3 N_c$.
For fixed number of colors $N_c$ and fixed number of flavors $N_f$, 
the theory flows in the IR to an isolated superconformal fixed point. It is possible however to define a systematic perturbative expansion
near the upper edge of the conformal window, taking the Veneziano limit $N_c \to \infty$, $N_f \to \infty$ with $N_f/N_c = 3 - \epsilon$.
The dilation operator can be evaluated order by order in $\epsilon$, and was indeed
completely determined to leading order (one loop) in \cite{Liendo:2011wc} following \cite{Poland:2011kg}.
Similarly one can set up an expansion for the dilation operator of the magnetic Seiberg-dual theory, near the lower edge of the conformal window, with $N_f / N_c = \frac{3}{2} + \tilde \epsilon$.
Seiberg duality implies that  the resummation of the $\epsilon$ expansion in the electric theory must coincide with the resummation
of the $\tilde \epsilon$ expansion in the magnetic theory. In the $SU(2,1|1)$ sector, the dilation operator is the same as in ${\cal N}=4$ SYM,
and thus obviously integrable, up to two loops in both expansions. The optimistic scenario
is  for the sector to remain integrable throughout the conformal window. It will be interesting to perform higher order checks in both $\epsilon$ and $\tilde \epsilon$.
Integrability would offer the exciting prospect of much more quantitative tests of Seiberg duality than presently possible.

Finally, one may even consider purely bosonic conformal gauge theories, and hope for integrability of the  $SU(2,1)$ sector,
\be \label{SU(2,1)sector}
SU(2,1)\mbox{ sector:}\qquad\qquad
(\Dm_{+ \ad})^n \,  \Fm_{+ +}\,  .
\ee
Only isolated fixed points are known for non-supersymmetric theories in four dimensions.
The simplest and most interesting case is QCD itself, in the Veneziano limit near the upper edge of the conformal window, $N_f/N_c =  11/2 -\epsilon$.
To leading order in $\epsilon$ (one loop) the dilation operator in the $SU(2,1)$ sector is trivially the same as in ${\cal N}=4$ SYM, 
but unlike the supersymmetric cases, we are not aware of a diagrammatic argument that this agreement should persist to two loops.
It would be very interesting to perform an explicit two-loop calculation and check integrability. 

If our  ${\cal N}=1$ and ${\cal N}=0$ speculations 
 turn out to be valid, at least in some models,
it will be because the  integrability structures of ${\cal N}=4$ SYM, while generically broken,
are  sufficiently robust to survive deformations and RG flows in the special universal sectors that we have isolated. On the dual string side (when available)
these sectors are captured entirely by the $AdS_5$ factor of the sigma model.
Our conjectures may be phrased as ``best case scenarios''. It will be worth investigating them
further.

\section*{Acknowledgments}
It is a pleasure to thank N. Gromov, 
V. Kazakov,
 J. Minahan, E. Pomoni,
  C. Sieg, M. Staudacher, P. Vieira and K. Zarembo for useful discussions and comments.
 This work is partially supported by the NSF under Grants PHY-0969919 and PHY-0969739. Any opinions, findings, and conclusions
or recommendations expressed in this material are those of the authors and do not necessarily
reflect the views of the National Science Foundation.

\newpage
\bibliographystyle{JHEP}
\bibliography{Completebbl}


\end{document}